# The Persistent Effects of Peru's Mining MITA: Double Machine Learning Approach

Master of Quantitative Economics

University of California, Los Angeles

Alper D. Karakas

Faculty Advisor: Prof. Denis Chetverikov

December 6, 2024

## Table of Contents

# 1. Abstract

This study examines the long-term economic impact of the colonial Mita system in Peru, building on Melissa Dell's foundational work on the enduring effects of forced labor institutions. The Mita, imposed by the Spanish colonial authorities from 1573 to 1812, required indigenous communities within a designated boundary to supply labor to mines, primarily near Potosí. Dell's original analysis uses a regression discontinuity design (RDD), leveraging the Mita boundary to estimate the Mita's legacy on modern economic outcomes. Her findings indicate that regions subjected to the Mita exhibit lower household consumption levels and higher rates of child stunting, suggesting an economic disadvantage.

In this paper, I replicate Dell's results, confirming her estimates of the Mita's negative impact. Extending this analysis, I apply Double Machine Learning (DML) methods—the Partially Linear Regression (PLR) model and the Interactive Regression Model (IRM)—to further investigate the Mita's effects. DML allows for the inclusion of high-dimensional covariates and enables more flexible, non-linear modeling of treatment effects, potentially capturing complex relationships that a polynomial-based approach may overlook. While the PLR model provides some additional flexibility, the IRM model allows for fully heterogeneous treatment effects, offering a nuanced perspective on how the Mita's impact varies across regions and district characteristics.

My findings suggest that the economic legacy of the Mita is both more substantial and spatially heterogeneous than originally estimated. The IRM results reveal that proximity to Potosí and other district-specific factors intensify the Mita's adverse impact, suggesting a deeper persistence of regional economic inequality. These findings underscore the value of using machine learning to address the realistic non-linearity present in complex, real-world systems, such as the legacy of the Mita. By modeling hypothetical counterfactuals more accurately, machine learning enhances my ability to estimate the true causal impact of historical interventions.

## 2. Introduction

The economic impact of historical institutions remains a fundamental question in economic history and development economics. In her influential work, Melissa Dell examines the long-term effects of the Mita system, a forced labor regime imposed by Spanish colonial authorities in Peru and Bolivia between 1573 and 1812. The Mita required indigenous communities within a specified geographic boundary to send a portion of their male population to work in silver and mercury mines, primarily centered around the city of Potosí. Dell's study provides evidence that regions historically subjected to the Mita experience lower household consumption levels and higher rates of stunted growth among children, suggesting a persistent legacy of economic disadvantage.

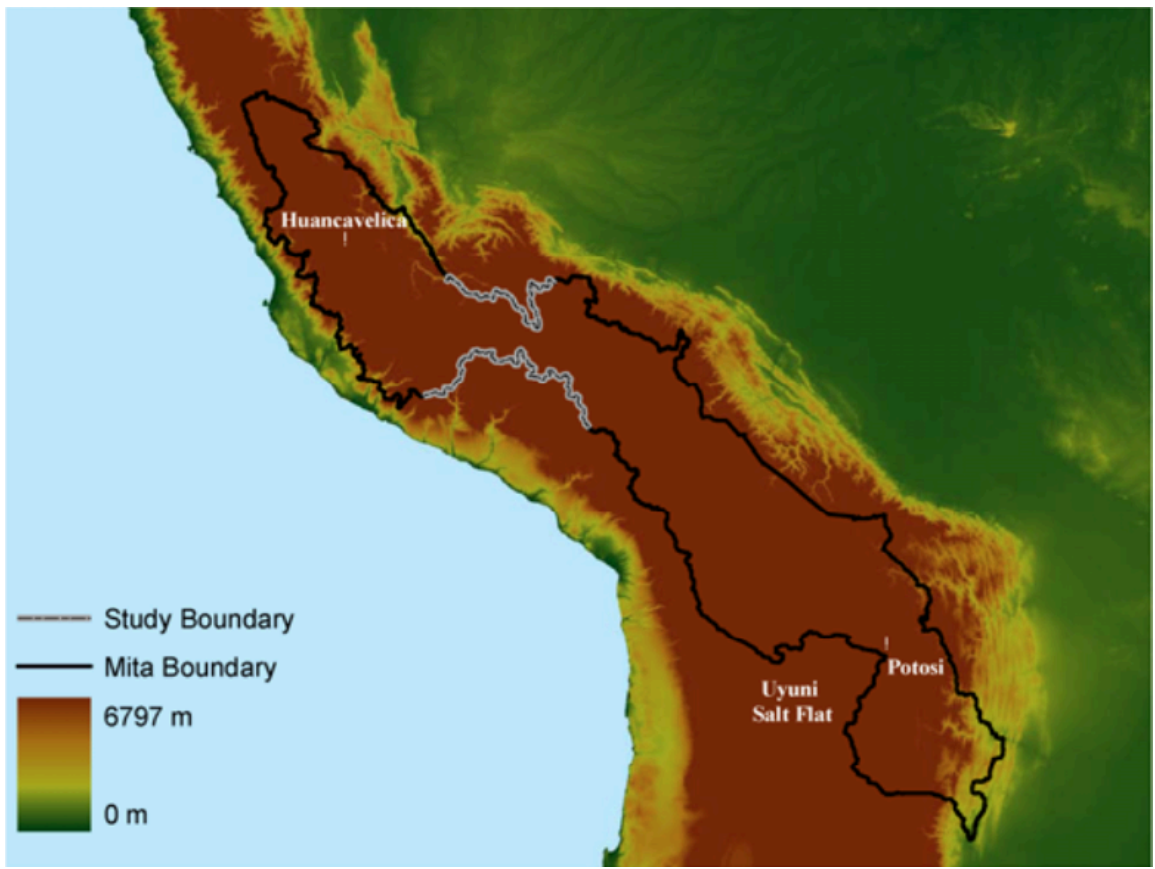

Dell's analysis employs a regression discontinuity design (RDD), leveraging the Mita boundary as a natural experiment to compare outcomes between districts within and outside the mandated labor region. Using geographic controls such as latitude, longitude, and distance to Potosí, Dell's study finds that districts exposed to the Mita have endured significant, negative economic effects, underscoring the lasting impact of colonial institutions. However, her approach relies on polynomial modeling, which may limit the flexibility needed to capture the complex, heterogeneous effects of the Mita on economic outcomes.

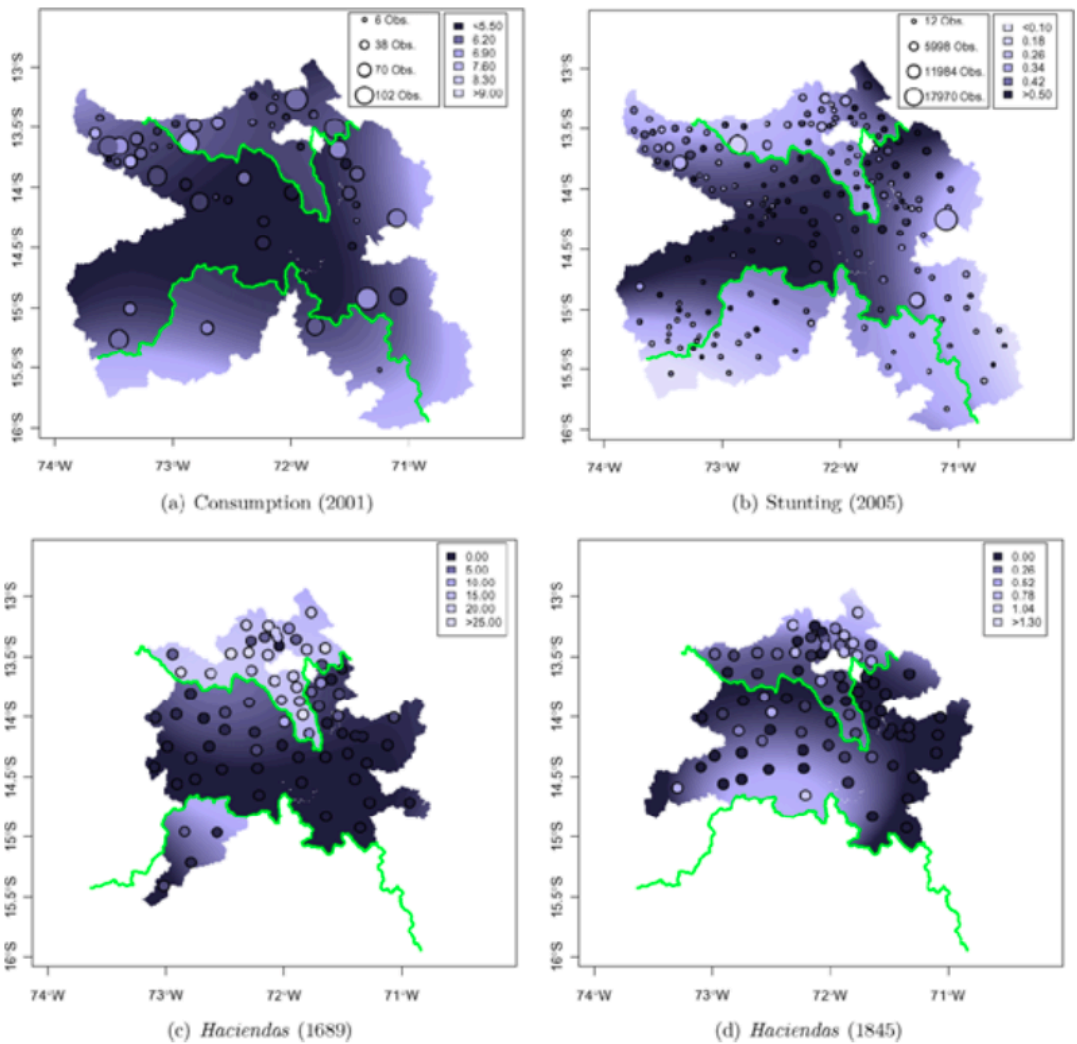

(a) Consumption (2001)  (b) Stunting (2005)

(c) *Haciendas* (1689)  (d) *Haciendas* (1845)

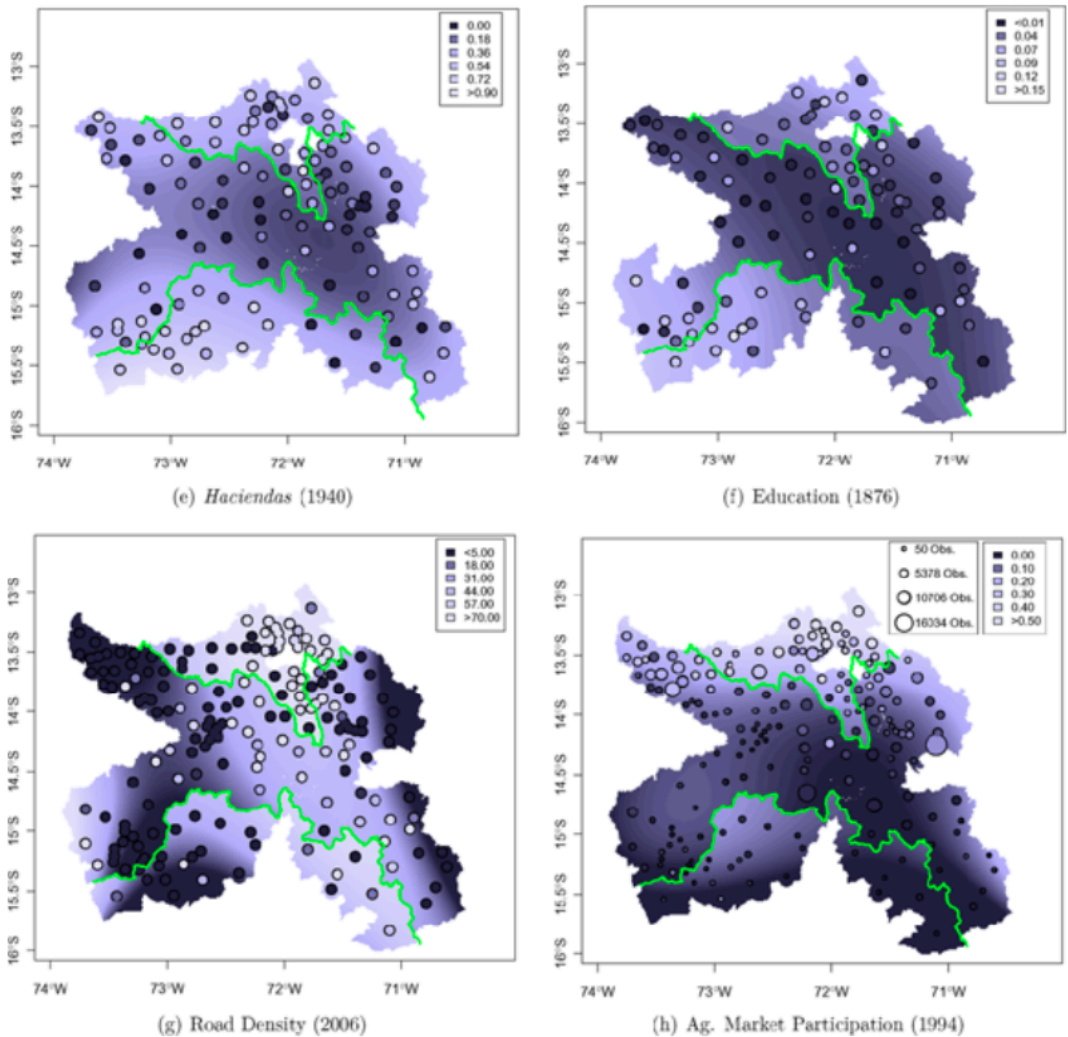

(e) *Haciendas* (1940)  (f) Education (1876)

(g) Road Density (2006)  (h) Ag. Market Participation (1994)

In this paper, I extend Dell's analysis by employing Double Machine Learning (DML) methods, specifically the Partially Linear Regression (PLR) model and the Interactive Regression Model (IRM), to re-estimate the economic legacy of the Mita system. DML allows for the inclusion of high-dimensional covariates and non-parametric interactions, providing a more flexible approach that better approximates counterfactual scenarios. By incorporating these advanced techniques, my analysis aims to provide a refined understanding of the Mita's legacy, investigating whether more flexible models yield stronger evidence of historical institutional persistence in modern economic outcomes.

## 3. Data

This analysis uses the same dataset as in Melissa Dell's study, *The Persistent Effects of Peru's Mining Mita*. The dataset comprises historical and contemporary socioeconomic information from various districts in Peru, focusing on the regions affected by the colonial mining Mita system and comparable regions outside the Mita boundary. The Mita was a forced labor system imposed by the Spanish colonial administration, requiring indigenous communities to send a proportion of their male population to work in the mines. The goal of this dataset is to allow for a comparison between areas that were part of the Mita and those that were not, thus providing a way to assess the Mita's long-term impact on socioeconomic outcomes.

Table I
Summary Statistics

|  | Dummy (Household Contributes to Mita) | Longitude | Latitude | Distance to Potosí |
|---|---|---|---|---|
| Count | 1478 | 1478 | 1478 | 1478 |
| Mean | 0.752368 | -0.334883 | -0.053991 | 8.963985 |
| Standard Dev, | 0.431783 | 1.203266 | 0.820277 | 1.449563 |
| Minimum | 0 | -2.30323 | -1.147286 | 6.231187 |
| 25% | 1 | -1.11573 | -0.751446 | 8.132674 |
| 50% | 1 | -0.22184 | -0.351446 | 9.316523 |
| 75% | 1 | 0.751221 | 0.786054 | 10.17482 |
| Maximum | 1 | 1.73011 | 1.436054 | 11.11004 |

(Continues)

Key variables in the dataset include both historical geographic measures and modern indicators of economic well-being. The primary treatment variable in this analysis is a binary indicator representing whether a district was part of the historical Mita system. Districts within the boundary of the Mita region are coded as treated, while districts outside the boundary serve as a control group. This variable allows for a clear comparison between Mita and non-Mita regions.

The dataset also includes geographic variables essential to the regression discontinuity design used in this analysis. These geographic controls are measured by a "running variable," which represents the location of each district in relation to the Mita boundary.

Table I—Continued

|  | Distance to Mita Boundary | Elevation (100m) | Mean Slope | Number of Infants | Number of Children |
|---|---|---|---|---|---|
| Count | 1478 | 1478 | 1478 | 1478 | 1478 |
| Mean | 40.639062 | 3.840895 | 7.129698 | 0.50203 | 1.217862 |
| Standard Dev, | 28.62655 | 0.378298 | 4.1237 | 0.730186 | 1.321108 |
| Minimum | 4.246131 | 2.44923 | 0.305039 | 0 | 0 |
| 25% | 18.29636 | 3.68349 | 3.6497 | 0 | 0 |
| 50% | 30.16593 | 3.88091 | 6.89444 | 0 | 1 |
| 75% | 67.47355 | 4.11275 | 9.94351 | 1 | 2 |
| Maximum | 98.50797 | 4.64621 | 17.5783 | 3 | 7 |

(Continues)

There are three key ways that location is measured: latitude and longitude coordinates, distance to the city of Potosí (where much of the Mita labor was directed), and the direct distance to the Mita boundary. These variables are integral to the identification strategy because they allow the analysis to compare districts that are geographically close to the Mita boundary, thereby strengthening the assumption that conditions across the boundary would have been similar if not for the Mita.

Table I—Continued

|  | Number of Adults | Boundary Segment 1 (Fixed Effect) | Boundary Segment 2 (Fixed Effect) |
|---|---|---|---|
| Count | 1478 | 1478 | 1478 |
| Mean | 2.536536 | 0.085927 | 0.288904 |
| Standard Dev, | 1.255328 | 0.280351 | 0.453407 |
| Minimum | 1 | 0 | 0 |
| 25% | 2 | 0 | 0 |
| 50% | 2 | 0 | 0 |
| 75% | 3 | 0 | 1 |
| Maximum | 10 | 1 | 1 |

(Continues)

In addition to geographic variables, the dataset contains contemporary economic indicators that serve as outcome variables in the analysis. These include measures of household consumption, child growth stunting rates, access to infrastructure, and reliance on subsistence agriculture. These outcome variables allow the study to capture the long-term economic effects of the Mita system by comparing socioeconomic conditions across Mita and non-Mita regions. The dataset's detailed demographic and economic data provide a robust basis for examining how historical institutional differences have persisted over time.

Table I—Continued

|  | Boundary Segment 3 (Fixed Effect) | Log Equiv. Household Consumption (2001) |
|---|---|---|
| Count | 1478 | 1478 |
| Mean | 0.384303 | 5.877284 |
| Standard Dev, | 0.486595 | 1.010021 |
| Minimum | 0 | 0 |
| 25% | 0 | 5.396283 |
| 50% | 0 | 5.916527 |
| 75% | 1 | 6.447267 |
| Maximum | 1 | 8.748576 |

By using Dell's original dataset, this analysis not only replicates her findings but also applies Double Machine Learning techniques—specifically the Partially Linear Regression Model (PLR) and Interactive Regression Model (IRM)—to uncover additional insights into the long-term effects of the Mita system. The dataset's historical, geographic, and economic information enables a rigorous evaluation of institutional persistence. Geographic measures, such as latitude/longitude and distance to Potosí, facilitate comparison of regions near the Mita boundary, providing a robust basis for identifying potential disparities. Key economic outcomes, including household consumption, child stunting rates, and infrastructure access, serve as indicators of modern economic development, allowing this study to examine how colonial institutions may have contributed to regional inequalities that persist today.

## 4. Replicating Dell's Results

This analysis replicates the findings of Melissa Dell's study by employing a regression discontinuity (RD) design. This RD framework leverages the historical Mita boundary as a natural discontinuity to compare economic outcomes between districts within and outside of the Mita region. By focusing on areas close to the Mita boundary, the RD design assumes that, absent the Mita, districts near the boundary would have had similar economic conditions. This design strengthens causal inference by minimizing unobserved differences that could otherwise confound the effect of the Mita system on economic outcomes.

To implement the RD design, the analysis uses models with cubic polynomials in various geographic running variables, which serve as measures of location relative to the Mita boundary. The use of polynomials allows the models to flexibly control for geographic variation across the boundary, accounting for the potential non-linear relationship between proximity to the Mita boundary and the outcomes. Specifically, Panel A employs a cubic polynomial in latitude and longitude, treating geographic coordinates as continuous variables that capture location without relying on any central reference point. Panel B incorporates a cubic polynomial in the distance to Potosí, the central city to which Mita laborers were required to travel. This distance metric is significant as it captures the historical connection to the labor demands imposed by the Mita. Panel C uses a cubic polynomial in the direct distance to the Mita boundary, allowing for a straightforward comparison of proximity to the boundary itself, regardless of orientation to Potosí.

These polynomials are combined with geographic controls, boundary fixed effects, and demographic controls to account for other sources of variation. The unit of observation is the household, with the dependent variable being log equivalalised household consumption in 2001. The replication successfully produces results consistent with those in Dell's original paper, as shown in *Table II: Living Standards*. Across Panels A, B, and C, the coefficients for the Mita variable are consistently negative, indicating that household consumption is lower in districts that were part of the Mita system compared to non-Mita districts. Regarding Panel A (Latitude and Longitude), the Mita coefficient is negative in all columns, with values ranging from -0.216 to -0.331. However, these coefficients are not highly statistically significant, suggesting that while there is a negative association between the Mita and household consumption, this relationship may not be robust when latitude and longitude are used as the running variable. For Panel B (Distance to Potosí), the coefficients for the Mita variable are more strongly negative, ranging from -0.307 to -0.358, with all estimates reaching a higher level of statistical significance. This pattern indicates that using distance to Potosí captures more residual variance in the outcome variable, as proximity to this city was directly related to the intensity of Mita obligations.

The significance of these estimates supports the interpretation that districts closer to Potosí and within the Mita boundary experienced sustained negative impacts on household consumption. For Panel C (Distance to Mita Boundary), the Mita coefficients are also negative and statistically significant across all columns, with values ranging from -0.223 to -0.277. These results imply that proximity to the Mita boundary itself is a strong predictor of lower economic outcomes, even when not anchored to Potosí. The statistical significance in Panel C reinforces the impact of the Mita on household consumption.

The consistently negative coefficients for the Mita variable across all models reflect the long-term economic consequences of the forced labor institution. According to Dell's analysis, "districts that were part of the Mita exhibit substantially lower levels of household consumption in 2001." The magnitude of these effects is notable: in some specifications, being within the Mita region is associated with a reduction in household consumption by more than 30 percent compared to comparable non-Mita areas. This sustained economic disparity suggests that the Mita's legacy has hindered economic development over generations, likely due to lower investments in infrastructure, human capital, and public

## Table II
## Living Standards

|  | Dependent Variable | | |
|---|---|---|---|
|  | Log Equiv. Household Consumption (2001) | | |
| Sample Within: | < 100 km of Bound | < 75 km of Bound | < 50 km of Bound |
|  | (1) | (2) | (3) |
| *Panel A. Cubic Polynomial in Latitude and Longitude* | | | |
| Mita | -0.2841 | -0.2164 | -0.3311 |
|  | (0.199) | (0.207) | (0.219) |
| R-squared | 0.059 | 0.060 | 0.069 |
| *Panel B. Cubic Polynomial in Distance to Potosí* | | | |
| Mita | -0.3368*** | -0.3070*** | -0.3286*** |
|  | (0.087) | (0.101) | (0.096) |
| R-squared | 0.046 | 0.036 | 0.047 |
| *Panel C. Cubic Polynomial in Distance in Mita Boundary* | | | |
| Mita | -0.277*** | -0.2300** | -0.2235** |
|  | (0.078) | (0.089) | (0.092) |
| R-squared | 0.044 | 0.042 | 0.040 |
| Geo. controls | yes | yes | yes |
| Boundary F.E.s | yes | yes | yes |
| Clusters | 71 | 60 | 52 |
| Observations | 1478 | 1161 | 1013 |

Notes: The unit of observation is the household in columns 1–3. Robust standard errors, adjusted for clustering by district, are in parentheses. The dependent variable is log equivalent household consumption (ENAHO (2001)) in columns 1–3. Mita is an indicator equal to 1 if the household's district contributed to the mita and equal to 0 otherwise (Saignes (1984), Amat y Juniet (1947, pp. 249, 284)). Panel A includes a cubic polynomial in the latitude and longitude of the observation's district capital, panel B includes a cubic polynomial in Euclidean distance from the observation's district capital to Potosí, and panel C includes a cubic polynomial in Euclidean distance to the nearest point on the mita boundary. All regressions include controls for elevation and slope, as well as boundary segment fixed effects (F.E.s). Columns 1–3 include demographic controls for the number of infants, children, and adults in the household. In column 1, the sample includes observations whose district capitals are located within 100 km of the mita boundary, and this threshold is reduced to 75 and 50 km in the succeeding columns. 78% of the observations are in mita districts in column 1, 71% in column 2, and 68% in column 3. Coefficients that are significantly different from zero are denoted by the following system: *10%, **5%, and ***1%.

goods in Mita districts. The results imply that the Mita system's forced labor requirements created persistent economic disadvantages that continue to affect living standards in these regions.

Dell's use of geographic variation in distance to Potosí and proximity to the Mita boundary highlights how the intensity of forced labor obligations has shaped long-term economic outcomes. Districts closer to Potosí, and therefore more affected by Mita labor demands, demonstrate lower household consumption, signifying deeper and more enduring impacts. Similarly, the robustness of the results when using the distance to the Mita boundary (Panel C) underscores the significance of physical proximity to the forced labor region itself in perpetuating these economic disparities.

In summary, the replicated results support Dell's conclusion that the Mita system has had lasting adverse effects on economic well-being, with substantial reductions in household consumption in regions subjected to the institution. This analysis, by replicating these results, confirms the robustness of Dell's findings and provides a basis for further exploration of institutional persistence and its influence on regional inequalities.

## 5. Double Machine Learning Approach

In this analysis, I apply Double Machine Learning (DML) to re-evaluate the findings of Melissa Dell's study on the economic effects of the Mita system. By using DML, I aim to capture complex, non-linear relationships in the data more effectively than traditional econometric methods, particularly the cubic polynomial specifications used in Dell's original study. The DML framework allows me to estimate counterfactual scenarios—such as what outcomes would look like if treated districts had not been part of the Mita and vice versa—mimicking the rigor of randomized controlled trials (RCTs) within an observational setting. This approach provides a refined, data-driven method to investigate the long-term impacts of the Mita on economic outcomes.

For each Double Machine Learning model, corresponding to each combination of column and panel in *Table II* (e.g., column 2, Panel B), I trained and optimized Neural Networks to serve as the prediction models. Neural Networks are particularly suited for this analysis because of their ability to flexibly approximate complex, high-dimensional relationships, handling non-linearities in ways that traditional cubic polynomials may not capture. By employing these models, I can more accurately balance treated and untreated groups across a broader range of covariates, allowing for a more precise estimation of the Mita's effect on household consumption.

DML is particularly powerful in settings with high-dimensional controls, as it incorporates a large number of covariates to control for factors that may affect both treatment assignment and economic outcomes. In this analysis, the Neural Networks provide a non-parametric, machine learning approach that enables the DML framework to handle complex interactions within the data, going beyond the polynomial models originally used by Dell. This non-linear flexibility is especially important for capturing potential geographic and socioeconomic variations that influence economic outcomes across districts.

A core feature of DML is its reliance on Neyman Orthogonality, which ensures that small errors in estimating nuisance parameters (i.e., the control functions for the outcome and treatment models) do not bias the treatment effect estimate. Neyman Orthogonality is crucial for maintaining robustness, as it allows the Neural Networks to model complex relationships without the risk of overfitting propagating into the treatment effect estimation. This property further strengthens the validity of my causal estimates, making DML well-suited for high-dimensional, machine learning-based analyses like this one.

By using Double Machine Learning with Neural Networks to reanalyze Dell's results, I introduce a more sophisticated, non-parametric approach to studying the economic legacy of the Mita system. This framework allows me to revisit Dell's findings with advanced techniques that better approximate counterfactual scenarios, enabling a deeper investigation into how historical institutions have contributed to persistent regional disparities in economic outcomes.

## 5.1 Double Machine Learning: Partially Linear Regression Model (PLR)

The Partially Linear Regression (PLR) model within the Double Machine Learning framework provides a flexible approach that combines traditional regression with machine learning to handle high-dimensional covariates. In the PLR model, the outcome variable $Y$ is expressed as:

$$Y = D\theta_o + g_o(X) + \zeta$$

where $Y$ represents log equivalised household consumption, $D$ is the treatment indicator (1 if a district was subject to the Mita, 0 otherwise), $\theta_o$ is the average treatment effect of the Mita, and $g_o(X)$ is an unknown function that captures the potentially non-linear effects of a high-dimensional vector of covariates $X$. Here, $\zeta$ represents a stochastic error term, with the assumption that $E(\zeta \mid D, X) = 0$. The treatment assignment $D$ is also modeled in relation to $X$ as follows:

$$D = m_o(X) + V$$

where $m_o(X)$ is the propensity score (the probability of receiving the treatment given $X$), and $V$ is an error term with $E(V \mid X) = 0$. This setup allows for a flexible specification of the treatment assignment process, where $D$ depends on $X$, capturing any selection into treatment that might depend on observed covariates. The PLR model assumes that the treatment effect $\theta_o$ is additive and constant across observations, which simplifies the estimation process but may limit the model's ability to capture heterogeneity in treatment effects. Specifically, the PLR model aims to estimate the Average Treatment Effect (ATE) under the assumption that the treatment effect is the same for all observations, regardless of their covariate values. This means that the treatment effect, $\theta_o$, is the ratio between $\partial E[Y \mid D, X]$ and $\partial D$.

This derivative represents the expected change in *Y* for a unit change in *D*, holding *X* constant. By estimating $g_o(X)$ non-parametrically (e.g., with machine learning techniques such as neural networks), the PLR model can flexibly control for high-dimensional covariates, allowing *X* to affect *Y* in complex ways. However, unlike the IRM model, the PLR approach constrains the effect of *D* to be additive, which may miss interactive or heterogeneous effects that vary across different values of *X*.

This limitation of the PLR approach means that while it provides a substantial improvement over traditional polynomial regression methods (such as those used in Dell's original analysis), it may not fully capture the Mita's varied effects across different regions and characteristics. By modeling $g_o(X)$ non-parametrically, the PLR model improves upon rigid polynomial specifications, but it is still less flexible than the fully interactive IRM model.

The results from the PLR model, as presented in *Table III*, reveal some important insights and differences compared to Dell's original findings. In line with Dell's results, the coefficients for the Mita variable remain negative across all specifications, indicating a persistent negative impact of the Mita system on household consumption in 2001. However, the statistical significance and magnitude of these effects vary across different panels and columns.

Regarding, Panel A (Latitude and Longitude) The PLR model yields negative Mita coefficients in all columns, but none are statistically significant. This result contrasts with Dell's findings, where there was at least a weak negative association using latitude and longitude as the running variable. The PLR model's lack of significance here suggests that the original polynomial model might have overemphasized the effect in these coordinates, whereas the more flexible PLR approach does not find strong evidence of an impact when using latitude and longitude alone.

## Table III
### Living Standards—PLR Approach

| | Dependent Variable | | |
|---|---|---|---|
| | Log Equiv. Household Consumption (2001) | | |
| Sample Within: | < 100 km of Bound | < 75 km of Bound | < 50 km of Bound |
| | (1) | (2) | (3) |

**Panel A. Cubic Polynomial in Latitude and Longitude**

| | | | |
|---|---|---|---|
| Mita | -0.0583 | -0.0379 | -0.0999 |
| | (0.078) | (0.096) | (0.108) |

**Panel B. Cubic Polynomial in Distance to Potosí**

| | | | |
|---|---|---|---|
| Mita | -0.1969*** | -0.0724 | -0.3593*** |
| | (0.054) | (0.066) | (0.067) |

**Panel C. Cubic Polynomial in Distance in *Mita* Boundary**

| | | | |
|---|---|---|---|
| Mita | -0.0816 | -0.0046 | -0.0823 |
| | (0.082) | (0.064) | (0.059) |
| Geo. controls | yes | yes | yes |
| Boundary F.E.s | yes | yes | yes |
| Observations | 1478 | 1161 | 1013 |

Notes: The unit of observation is the household in columns 1–3. Robust standard errors are in parentheses. The dependent variable is log equivalent household consumption (ENAHO (2001)) in columns 1–3. Mita is an indicator equal to 1 if the household's district contributed to the mita and equal to 0 otherwise (Saignes (1984), Amat y Juniet (1947, pp. 249, 284)). Panel A includes a cubic polynomial in the latitude and longitude of the observation's district capital, panel B includes a cubic polynomial in Euclidean distance from the observation's district capital to Potosí, and panel C includes a cubic polynomial in Euclidean distance to the nearest point on the mita boundary. All models include controls for elevation and slope, as well as boundary segment fixed effects (F.E.s). Columns 1–3 include demographic controls for the number of infants, children, and adults in the household. In column 1, the sample includes observations whose district capitals are located within 100 km of the mita boundary, and this threshold is reduced to 75 and 50 km in the succeeding columns. 78% of the observations are in mita districts in column 1, 71% in column 2, and 68% in column 3. Coefficients that are significantly different from zero are denoted by the following system: *10%, **5%, and ***1%.

The results in Panel B (Distance to Potosí) align more closely with Dell's original findings, showing a significant negative effect of the Mita in columns (1) and (3). The coefficient values are similar to those in Dell's study: proximity to Potosí remains a strong predictor of lower household consumption in Mita regions. This pattern supports the historical context of Potosí's central role in the Mita system, with areas closer to the city experiencing more intense and prolonged economic consequences. However, the PLR model further emphasizes this effect's robustness, especially with the stronger significance level in column (1), reinforcing the idea that Potosí's proximity is crucial for understanding the Mita's impact. In Panel C (Distance to Mita Boundary), the PLR model's results show weak evidence of a negative Mita effect, with coefficients close to zero and no statistical significance across columns. This differs from Dell's findings, where proximity to the Mita boundary showed a statistically significant impact. The PLR model's results suggest that when a more flexible, machine-learning approach is used, the effect of distance to the Mita boundary may be less pronounced than originally indicated. This may imply that distance to Potosí (as seen in Panel B) is a more accurate predictor of the Mita's long-term economic impact than mere proximity to the Mita boundary.

Overall, the PLR model tells a slightly different story than Dell's original findings. While it confirms the negative economic legacy of the Mita system, it nuances the interpretation of geographic influence. The PLR results suggest that proximity to Potosí plays a more significant role in explaining economic disparities than latitude/longitude or distance to the Mita boundary alone. This distinction highlights the importance of selecting appropriate geographic measures in historical analyses and suggests that the intensity of exposure to the Mita, rather than simple proximity, is a key determinant of its lasting effects on economic outcomes.

## 5.2 Double Machine Learning: Interactive Regression Model (IRM)

The Interactive Regression Model (IRM) within the Double Machine Learning framework provides a powerful approach for estimating treatment effects when those effects are fully heterogeneous, meaning they vary across different groups or conditions. Unlike the Partially Linear Regression (PLR) model, which assumes a linear and additive relationship, the IRM model allows the treatment effect to interact with covariates, capturing more complex, non-linear relationships that may vary significantly across different districts. The IRM model takes the following form:

$$Y = g_0(D, X) + U$$

where $Y$ is the outcome variable (log equivalised household consumption), $D$ is the binary treatment variable indicating Mita participation (with $D \in [0, 1]$), and $X$ is a high-dimensional vector of covariates. Here, $g_0(D, X)$ is a function that allows the effect of $D$ on $Y$ to vary depending on the covariates $X$, while $U$ represents the stochastic error term, satisfying $E(U \mid X, D) = 0$. This formulation permits fully interactive effects between $D$ and $X$, making it a suitable approach for understanding heterogeneous treatment effects.

In addition to modeling the outcome, the IRM approach also models the treatment assignment $D$ as:

$$D = m_0(X) + V$$

where $m_0(X)$ is the probability of treatment (the propensity score), and $V$ is an error term such that $E(V \mid X) = 0$. This specification allows the treatment assignment to depend on the covariates $X$ while ensuring that residual noise in treatment assignment does not bias the estimation of treatment effects.

The IRM model aims to estimate two key parameters: the Average Treatment Effect (ATE) and the Average Treatment Effect on the Treated (ATTE). These parameters are defined as follows:

$$\text{Average Treatment Effect (ATE)} = \theta_0 = E[g_0(1, X) - g_0(0, X)]$$

This represents the expected difference in outcomes if all districts were treated versus if none were treated, averaging over all values of $X$.

$$\text{Average Treatment Effect on the Treated (ATTE)} = \theta_0 = E[g_0(1, X) - g_0(0, X) \mid D = 1]$$

This captures the treatment effect specifically for districts that were part of the Mita, providing insight into the impact of treatment on the treated subset of the population.

The IRM model is especially well-suited for this analysis because it allows $g_0(D,X)$ to vary with $X$, capturing the non-linear and potentially complex relationships between covariates and outcomes that may have been introduced by the historical Mita system. In contrast, the PLR model assumes an additive treatment effect of the form $Y = D\theta_0 + g_0(X) + \zeta$, where $D$ has a linear effect on $Y$. This additive form limits PLR's capacity to account for interactive effects and fully heterogeneous responses to treatment, potentially oversimplifying the Mita's impact. By modeling these interactive, non-linear effects directly, IRM provides a more flexible and accurate approach for estimating the true economic legacy of the Mita system. The interaction between $D$ and $X$ in the IRM model allows it to capture how the Mita's impact varies across regions and population characteristics, offering a closer approximation to the true effect than either the cubic polynomial models in Dell's original analysis or the PLR approach.

The IRM results, shown in *Table IV*, indicate significant and consistently negative effects of the Mita across all panels and columns. These results provide a more robust and nuanced picture of the Mita's economic impact than both Dell's original findings and the PLR results. Regarding Panel A (Latitude and Longitude), the IRM model yields strongly negative and statistically significant Mita coefficients across all columns, ranging from -0.9131 to -0.1346.

### Table IV
### Living Standards—IRM Approach

| Sample Within: | Dependent Variable | | |
|---|---|---|---|
| | Log Equiv. Household Consumption (2001) | | |
| | < 100 km of Bound | < 75 km of Bound | < 50 km of Bound |
| | (1) | (2) | (3) |
| Panel A. Cubic Polynomial in Latitude and Longitude | | | |
| Mita | -0.1346*** | -0.9131*** | -0.6709*** |
| | (0.033) | (0.049) | (0.056) |
| Panel B. Cubic Polynomial in Distance to Potosí | | | |
| Mita | -0.4614*** | -0.4419*** | -0.4039*** |
| | (0.036) | (0.037) | (0.038) |
| Panel C. Cubic Polynomial in Distance in Mita Boundary | | | |
| Mita | -0.6363*** | -0.1946*** | -0.1268** |
| | (0.053) | (0.046) | (0.056) |
| Geo. controls | yes | yes | yes |
| Boundary F.E.s | yes | yes | yes |
| Observations | 1478 | 1161 | 1013 |

Notes: The unit of observation is the household in columns 1–3. Robust standard errors are in parentheses. The dependent variable is log equivalent household consumption (ENAHO (2001)) in columns 1–3. Mita is an indicator equal to 1 if the household's district contributed to the mita and equal to 0 otherwise (Saignes (1984), Amat y Juniet (1947, pp. 249, 284)). Panel A includes a cubic polynomial in the latitude and longitude of the observation's district capital, panel B includes a cubic polynomial in Euclidean distance from the observation's district capital to Potosí, and panel C includes a cubic polynomial in Euclidean distance to the nearest point on the mita boundary. All models include controls for elevation and slope, as well as boundary segment fixed effects (F.E.s). Columns 1–3 include demographic controls for the number of infants, children, and adults in the household. In column 1, the sample includes observations whose district capitals are located within 100 km of the mita boundary, and this threshold is reduced to 75 and 50 km in the succeeding columns. 78% of the observations are in mita districts in column 1, 71% in column 2, and 68% in column 3. Coefficients that are significantly different from zero are denoted by the following system: *10%, **5%, and ***1%.

These coefficients are both larger and more significant than those in the PLR model, suggesting that the interactive structure of IRM captures a more pronounced negative impact of the Mita on household consumption, particularly when considering geographic coordinates. Compared to Dell's original results, the IRM model indicates a stronger effect, highlighting that Dell's polynomial approach may have underestimated the impact when using latitude and longitude as running variables.

In Panel B (Distance to Potosí), the IRM results show consistently negative and highly significant effects, with coefficients between -0.4039 and -0.4614 across all columns. These findings align with Dell's original results, which also highlighted the significant negative impact of proximity to Potosí, but the IRM model further strengthens this relationship by allowing the effect to vary interactively. Compared to the PLR approach, which also found significant results in this panel, the IRM model shows slightly stronger coefficients, suggesting that the interactive effects capture additional nuances related to distance and other covariates. Lastly, the IRM results in Panel C (Distance to Mita Boundary) reveal significant negative effects across all columns, with coefficients ranging from -0.1268 to -0.6363. This contrasts with the PLR results, where coefficients were close to zero and not statistically significant.

The IRM model shows that proximity to the Mita boundary has a meaningful negative impact when the treatment effect is allowed to vary interactively with district characteristics, which the PLR model may have missed by assuming additivity. The IRM results provide a refined and detailed understanding of the Mita system's economic legacy. Compared to both Dell's original findings and the PLR results, the IRM model reveals a more substantial and heterogeneous impact, suggesting that the Mita's influence on household consumption is both large and varies across regions. This indicates that the Mita's legacy is complex, with its effects depending on specific district characteristics like distance to Potosí and proximity to the Mita boundary. I assert that the true effect of the Mita system on economic outcomes is best captured by the IRM model. The IRM results suggest that the Mita's forced labor requirements have created lasting economic disadvantages that are heterogeneous, shaped by geographic and socioeconomic factors. Through the IRM approach, there is a clearer and more accurate picture of how historical institutions like the Mita have perpetuated regional economic disparities, thus providing a closer approximation to the true effects than either the cubic polynomial regression or the PLR model.

## 6. Conclusion

This report investigates the long-term economic effects of Peru's colonial Mita labor system on modern household consumption levels, building on Melissa Dell's original research. The central research question guiding my analysis is whether the Mita system has had a lasting, negative impact on economic outcomes and whether modern econometric techniques can deepen my understanding of this historical institution's legacy. While Dell's analysis used a regression discontinuity design with cubic polynomials to estimate these effects, I apply Double Machine Learning (DML) methods to investigate whether a more flexible, non-parametric approach reveals additional insights. My goal is to determine if using DML changes the estimated impact of the Mita and if it brings me closer to capturing the true effect.

To address this question, I first replicated Dell's findings, confirming that the Mita system is associated with lower household consumption levels in regions historically affected by the labor mandate. Next, I applied two DML approaches—the Partially Linear Regression (PLR) model and the Interactive Regression Model (IRM)—to estimate the treatment effect of the Mita with greater flexibility and accuracy. The DML framework allows me to control for high-dimensional covariates while leveraging machine learning models, providing a robust method to approximate counterfactual outcomes in a way that Dell's polynomial-based approach may not capture fully. The PLR model adds flexibility to Dell's framework by allowing the covariates' effects on consumption to be modeled non-parametrically, while the IRM model further allows the treatment effect to interact with these covariates, accounting for heterogeneous effects that vary by district characteristics.

My findings reveal that the IRM approach, in particular, offers a refined perspective on the Mita's impact, suggesting a stronger and more heterogeneous effect on household consumption than initially estimated by Dell. While Dell's results indicated a significant negative impact of the Mita, especially in regions close to Potosí, the DML analysis highlights that these effects are not uniform across regions. The IRM results show that the Mita's impact varies based on geographic and socioeconomic factors, with some districts experiencing more pronounced economic disadvantages than others. This nuanced understanding suggests that the Mita's legacy has created regional economic disparities that are more complex than a single average treatment effect can convey.

In conclusion, my DML-based analysis suggests that the true economic impact of the Mita system is both larger and more variable than Dell's original findings indicate. By allowing treatment effects to vary interactively with covariates, the IRM approach captures the heterogeneous nature of the Mita's influence on modern economic outcomes, revealing a deeper persistence of inequality. This research underscores the value of advanced econometric methods like DML for historical economic analysis, demonstrating how flexible, non-linear models can provide more accurate estimates of treatment effects in settings where historical and geographic factors interact in complex ways.